# Domain-Pair Intertwined Topological Domain Structure in Elemental Bi Monolayer

Yunfei Hong[1,2], Junkai Deng[1,*], Yang Yang[1], Ri He[3], Zhicheng Zhong[3], Xiangdong Ding[1], Jun Sun[1], Jefferson Zhe Liu[2,*]

[1]*State Key Laboratory for Mechanical Behavior of Materials, Xi'an Jiaotong University, Xi'an 710049, China*

[2]*Department of Mechanical Engineering, The University of Melbourne, Parkville, VIC 3010, Australia*

[3]*Key Laboratory of Magnetic Materials Devices & Zhejiang Province Key Laboratory of Magnetic Materials and Application Technology, Ningbo Institute of Materials Technology and Engineering, Chinese Academy of Sciences, Ningbo 315201, China*

[*]Corresponding author

E-mail: junkai.deng@mail.xjtu.edu.cn; zhe.liu@unimelb.edu.au

## Abstract

Ferroelectric domain structures, separated by domain walls, often display unconventional physics and hold significant potential for applications in nano-devices. Most naturally growth domain walls are charge-neutral to avoid increased electrostatic energy, while the intrinsically stable charged 180° domain walls in Bi monolayer challenged this conventional knowledge and emerged an unexplored field. Here, using machine-learning potential and molecular dynamics (MD) simulations, we investigated the finite-temperature dynamics of domain walls and discovered a domain-pair intertwined topological domain structure in Bi monolayer. In 180° domain walls, a unique polarization switching mechanism is observed, characterized by the out-of-plane shuffle of Bi atoms without bond breaking. This shuffle mechanism reverses the charge properties of Bi atoms, transforming Bi anions into cations and vice versa, ultimately reversing the polarization. Then, we observed a topological multi-domain structure with two groups of domain pairs intertwined. The charged 180° domain walls form local domain pairs, with the 90° domain walls emerge between different domain pairs. This multi-domain maintains a stable topological structure within the strain range ($\varepsilon_x$ = 0 to 4.70%) and exhibits rich domain wall reactions under further applied strain. Our findings provide insights into the charged 180° domain walls and the related topological domain structures, enabling new opportunities for applications in electronic and nano-electronic devices.

## Introduction

Ferroelectric domains are structural units in ferroelectric materials, characterized by uniform polarization orientations, separated by distinct domain walls, most commonly the 180° and 90° domain walls[1,2]. These domain walls display reduced dimensionality and altered symmetry compared to bulk materials, resulting in novel physical properties such as enhanced or reduced conductivity[3–5], electromagnetic coupling[6–9], and modified mechanical behavior[10]. Furthermore, theoretical and experimental studies have revealed that ferroelectric domains, under the interplay of elastic, electrostatic, and gradient energies, can give rise to complex ferroelectric topological domains, such as vortex domains[11–13], flux-closure domains[14,15], bubble domains[16,17], and skyrmions[18–20]. It is discovered that these naturally growth domain walls are predominantly charge-neutral, exhibiting head-to-tail configurations, such as the four head-to-tail 90° domain walls in flux-closure domains[15] and continuous local polarization rotation in vortex domains[13]. Due to the instability arising from increased electrostatic energy caused by charge accumulation, charged domain walls[1,21] are rare and typically associated with extrinsic defects[22–24] or external fields[25].

Recent experimental studies have challenged this conventional understanding by revealing the widespread presence of intrinsically stable charged 180° domain walls (head-to-head or tail-to-tail) in single-elemental Bi monolayer[26,27]. The ferroelectricity in Bi monolayer originates from the energy splitting of $p_z$ orbitals, which induces charge transfer between Bi atoms, resulting in two distinct chemical bonding states and the spontaneous polarization[28–30]. Further research indicates that the unique lone-pair activated polarization in elemental Bi monolayer makes strain energy the dominant factor in domain wall stability, rather than electrostatic energy[26]. This allows charged domain walls to exist intrinsically and stably in Bi monolayers. This stable charged domain structure goes beyond conventional knowledge and has emerged as an unexplored field, potentially offering novel physical properties and new topological domain configurations, with significant potential for applications in two-dimensional electronic devices.

In this work, using machine learning potential and molecular dynamics (MD) simulations, we investigated the finite-temperature dynamics of domain walls and discovered a unique domain-pair intertwined topological domain structure in Bi monolayer. Through continues heating-cooling simulations, a second-order ferroelectric phase transition was discovered in Bi monolayer. We explored the domain wall energies and mobilities of the four typical domain walls (charged and neutral 90° (or 180°) domain walls at finite temperatures. The difference of domain wall energy is relatively small, but the mobility of 180° domain walls is significantly higher than that of 90° domain walls at high temperature (300K). The atomic structure analysis of charged 180° domain wall reveals an out-of-plane shuffle (without bond breaking)in polarization reversal, which differs significantly from the bond-breaking behavior in 90° domain walls. Then, we discovered a domain-pair intertwined topological domain structure that exhibits the coexistence of 180° and 90° domain walls. When subjected to quasi-static tensile strain, this multi-domain reveals stable topological structure and rich domain wall reactions. Our findings provide insights into the ferroelectric phase transitions and unique behavior of charged domain walls in elemental ferroelectrics, highlighting the potential of strain engineering to manipulate domain structures and paving the way for new applications in electronic and nano-electronic devices.

## Results
### Crystal structure of Bi monolayer
Bi monolayer, one of the group-V (As, Sb, and Bi) elemental ferroelectric materials, crystallizes in an asymmetric washboard structure with space group *Pmn2₁* (Figure 1a)[27,28]. Due to the energy splitting of $p_z$ orbitals, Bi monolayer exhibits two distinct chemical environments, enabling charge transfer between Bi atoms and resulting in intrinsic electric polarization ($P_s$). As illustrated in Figure 1a, Bi monolayer forms a rectangular lattice ($b > a$) with intrinsic buckling ($d_z \neq 0$) that is strongly coupled to the polarization. In the configuration shown in Figure 1a, Bi3 and Bi4 have gained electrons, becoming anions, while Bi1 and Bi2 act as cations. Due to the coupling between strain and electric polarization, monolayer Bi exhibits four energetically equivalent polarization variants, each characterized by a distinct intrinsic polarization (Figure 1b). Previous studies have demonstrated that $d_z$ governs the charge transfer[29,31], and the reversal of $d_z$ can invert the charge transfer, enabling opposite polarization. To represent the local polarization, we define an atomistic order parameter: $\vec{R} = (\vec{R}_{Bi1} - \vec{R}_{Bi3} + \vec{R}_{Bi2} - \vec{R}_{Bi4}) \times d_z$. This is achieved by projecting the relative positions of the four Bi atoms onto their local tangent plane. The orientation of the atomistic order parameter ($\vec{R}$) is denoted using four colors, as shown in Figure 1b.

**Interatomic potential**

The machine learning potential for MD simulations was fitted to an *ab initio* database of 16546 configurations over a range of temperatures and pressures. This potential achieves the DFT-level accuracy, with a root mean square error of $RMSE_{energy}$ = 1.365 meV/atom and $RMSE_{force}$ = 0.586 meV/Å (Figure S1, Supporting Information). It has been shown to well reproduce the crystal structure, phonon spectrum, strain-energy curve, and domain switching process, which allows us to use it on larger cells and longer-time simulations (TABLE S1, Figure S2, Supporting Information).

**Temperature induced second-order ferroelectric phase transition**

To investigate the ferroelectric phase transition of Bi monolayer, we perform a continuous heating-cooling cycle, increasing the temperature from 200 K to 500 K and then decreasing it back to 200 K at the same rate of 1.5 K/ps. Figure 1c shows the evolution of potential energy (related to the chemical bonding) as a function of temperature during heating and cooling cycle, demonstrating good continuity. Similarly, the evolution of average lattice constant as a function of temperature also exhibits good continuity (Figure 1d). However, the curve of average lattice constant *b* reveals a kink around 350 K, where the slope of the curve (coefficient of thermal expansion) changed abruptly, suggesting a phase transition at this temperature. The temperature-polarization curve from the isothermal simulation at various temperatures confirms the ferroelectric phase transition in Bi monolayer (Figure 1e), with the Curie temperature ($T_c$) approximately 360 K. The stable configurations at different temperatures are illustrated in Figure S3, Supporting Information, illustrating the gradual disappearance of long-range polarization. The continuity of potential energy and lattice constant is consistent with the characteristics of a second-order phase transition. Here, we repeated the heating-cooling cycle simulations (Figure S4, Supporting Information), and all of them show the similar results in both potential energy and lattice constants, suggesting the reliability of our molecular dynamics (MD) simulation results.

The typical atomic configurations of Bi monolayer at the specific temperature (200 K, 340 K, and 500 K) are presented in Figure 1f. Each atomic configuration is a stable structure, obtained after an additional 200 ps of isothermal simulation. Starting with a single-domain Bi monolayer at 200 K, with the polarization of (0, $-P_s$), we gradually heated the system to 500 K at a rate of 1.5 K/ps. As the temperature rises to 340 K, a few nano-regions with different polarization orientations begin to emerge, while the majority of the regions still maintain the previous polarization. This leads to the reduction in total polarization. The isothermal simulation reveals that these nano-regions are not fixed while randomly triggered over time (Figure S5, Supporting Information). As the temperature rises to 500 K, the long-range polarization disappears, while local regions still exhibit virtual polarization, forming numerous polar nano-regions. The total polarization of the entire system, as well as the long-term average of polarization in local region is zero, indicating that the polarization in these local regions fluctuates over time. As a result, the average atomic structure of paraelectric phase is similar to black phosphorus (Figure 1e).

**Finite-temperature dynamics of four typical domain walls**
Figure 2a shows four typical types of domain walls, charged and neutral 90° (or 180°) domain walls (90c, 90n, 180n, and 180c), which have been theoretically and experimentally reported[26]. The 90° domains patterns are formed by mutually perpendicular energetically equivalent domain variants separated by domain walls, whereas the 180° domains consist of two opposite domain variants (0, $\pm P_s$). For charged 90° domain walls, the polarization configuration around the domain walls is head-to-head (or tail-to-tail), resulting in charge accumulation at the domain walls. In contrast, the polarization configuration of neutral 90° domain walls is head-to-tail, which prevents charge accumulation. Similarly, charged 180° domain walls also have a head-to-head (or tail-to-tail) polarization configuration, leading to charge accumulation. In contrast, for neutral 180° domain walls, the polarization is parallel to the domain walls, effectively preventing charge accumulation as well.

Taking advantage of our high-quality machine learning potential, we can investigate the finite-temperature dynamics of four typical domain walls using MD simulation. We calculated the domain walls energy and domain wall mobility at various temperatures (300 K, 200 K, and 100 K). For 2D systems, domain walls energy ($E_{DWs}$) is defined as $E_{DWs} = \frac{E_{total} - E_0}{L}$. Here, $E_{total}$ is the energy of the system, $E_0$ is the energy of the single domain, and $L$ represents the length of domain walls. Figure 2b summaries the domain walls energy results at 300 K, alongside a comparison with the DFT calculations at 0K (Figure S6, Supporting Information). The results show $E_{DWs}(90c) > E_{DWs}(90n) > E_{DWs}(180n) > E_{DWs}(180c)$. The consistent results were obtained at 100 K and 200 K (Figure S7, Supporting Information). It is consistent to the widespread presence of charged 180° domain walls observed in experiments[27]. It is worth noting that the domain energy differences are relatively small at a given temperature, implying that their coexistence in one sample.

We also evaluate the domain wall mobility using root mean square displacement[32,33] ($RMSD = \sqrt{\frac{1}{N}\sum_{i=1}^{N}|\Delta \vec{R}_i|^2}$). Here, $N$ represents the number of sample positions on the domain walls, which was set to ten. The vector $\Delta \vec{R}_i$ represents the displacement of the *i*-th sampling point on the domain wall relative to its previous position, indicating the movement of domain walls over time. As illustrated in Figure 2c, the 180° domain walls exhibit with higher mobility (RMSD = 15.67 Å, 22.75 Å) compared to the 90° domain walls (RMSD = 4.58 Å, 5.79 Å) at 300 K. Indeed, the 180° domain walls display more rough in the atomic structure (Figure 2a). As the temperature decreases to 200 K and 100 K, the mobility of 180° domain walls decreases significantly, reaching a value comparable to that of the 90° domain walls (Figure S8, Supporting Information). Our MD simulation show that the 180c domain walls exhibit increasingly structural irregularities in the domain walls with the temperature increases (Figure S9, Supporting Information). Eventually, at 400 K, it transitioned into a paraelectric phase.

## Shuffling polarization switching in 180° domain walls

The atomistic structure of charged 180° domain walls can be viewed in two typical regions (labeled 1 and 2 in Figure 2a) to provide insights into the polarization switching around domain walls. In Figure 2d, the circles represent Bi atoms and solid lines are chemical bonds. The red (High) and blue (Low) circles are used to distinguish two types of vertical Bi pairs atoms based on their $z$-direction distortions. In region 1, as shown in the side view of Figure 2d, the arrangement of vertical atom pairs follows the periodic sequence 'Low-High-Low-High' on the left side of the domain wall, but 'High-Low-High-Low' sequence on the right side. The transition across the boundary can be achieved by re-shuffle the sequence. Note that this shuffling operation does not change the bonding arrangement between atoms. The top view (only show top layer Bi atoms) provides a clearer clue of this transformation around the domain wall. The green and yellow boxes represent the unit cell on the opposite side of domain boundary. The corner Bi atoms has low $z$ coordinate and the central Bi has a high $z$ coordinate. These two unit cells are mirror images (with a slight shift) across the boundary. To help understanding the structural change, we draw a vertical dashed line crossing a colinear of Bi vertical pairs in Figure 2d. On the top part (0, $+P_s$), the central Bi on the dashed line has a high $z$ coordinate (red). When cross the boundary to the bottom part, the Bi on the line turns to be the corner Bi with low $z$ coordinate (blue), indicating that the vertical Bi pairs shuffle along $z$ direction. Meanwhile, the neighboring Bi vertical pairs also shuffle along $z$ direction. This out-of-plane shuffle does not break (or reform) chemical bonds (solid lines). Similar observations were noted in region 2. A quantitative analysis of the charged 180° domain structure was performed to confirm the shuffling polarization switching. In Figure 2e, the reversal of buckling ($d_z$), associated with the shuffle of Bi atoms in $z$ direction, dominates the reversal of polarization. The domain wall spans approximately six units, within which the buckling parameter ($d_z$) varies continuously and eventually reverses, corresponding to the gradual reversal of polarization. In contrast, the in-plane displacement parameter $d_y$ remains unchanged. Since bond-breaking domain switching would typically alter $d_y$, this invariance confirms that the domain switching is driven solely by the out-of-plane shuffle of Bi atoms. The energy barrier for the shuffle mechanism in Bi monolayer is 38.13 meV/unit, in contrast to bond-breaking mechanism is at 162.58 meV/unit (Figure S10, Supporting Information).

## Intertwined domain pairs and the resulted checkerboard multi-domain structure

Next, we explored the multi-domain structures of Bi monolayer at combined condition by constructing an initial structure with random polarizations and cooling it from 500 K to 300 K. Figure 3a illustrates the typical formation process toward the checkerboard domain structure. It starts from a randomly polarized structure at 500 K, where various polarizations are shown using the color map in Figure 1b. The color map distinguishes different polarization directions. The sample was cooled to 300 K at a rate of 3.33 K/ps. At 433 K (inset at 20 ps), many polar nano-regions emerge. These polar nano-regions fluctuate over time without the long-range polarization. Upon cooling below the transition temperature ($\approx 360$ K), the polar nano-regions compete and eventually merge with surrounding domains to form clear ferroelectric domains. As shown in the inset (60 ps instead at 300 K), the charged 180° domain walls are observed widely in the Bi monolayer, forming local head-to-head (or tail-to-tail) domain pairs separated by the 180c boundaries. There are two groups of domain pairs: yellow and green arrows ($y$-direction pairs), red and blue arrows ($x$-direction pairs). Interestingly, we observed the formation of 90° domain walls to separate them, leading to the intertwined domain pairs structure. During the isothermal simulation at 300 K, these ferroelectric domains continued to grow to minimize total domain wall area and system energy. Ultimately, a domain-pair intertwined checkerboard domain with coexisting 180° and 90° domain walls was obtained (inset at 200 ps). Additional details on the evolution of the checkerboard domain structure are provided in Figure S11, Supporting Information. To confirm its stability, isothermal simulations of checkerboard domain structure were performed at 100 K, 200 K, 300 K, and 400 K (Figure S12, Supporting Information).

To reduce thermal fluctuations, the multi-domain was maintained at 10 K for 200 ps to obtain a stable and clearer domain structure. As shown in Figure 3b, the multi-domain structure displays four equivalent domain variants represented by different colors, which can be grouped into two pairs. The dominant domain pairs are characterized by $(0, \pm P_s)$ polarizations (green and yellow), resulting in charged 180° domain walls. Meanwhile, vertically oriented nano-domain pairs with $(\pm P_s, 0)$ polarizations (blue and red) form between the dominant domain pairs. These two domain pairs are separated by 90° domain walls, leading to the coexistence of 180° and 90° domain walls. The region 1 is a representative area where the 180° and 90° domain walls intersect, featuring a cruciform structure. Figure 3c illustrates the atomistic structure of region 1. The 90° domain walls (90n, 90c) are highlighted in gray, showing the bond breaking in the domain wall area. The charged 180° domain walls are represented by red solid lines, resulting from out-of-plane shuffle without bond breaking. The charged 180° domain walls traverse the region from left to right and twice intersect with the 90° domain walls, forming two types of domain pairs. Along the charged 180° domain walls, the domain pairs on the left are the dominate domain pairs (green and yellow). After the first intersection with 90° domain walls, these domain pairs transform into nano-domain pairs (blue and red) and then revert to the dominate domain pairs (green and yellow) after the second intersection. The local head-to-head configuration around the charged 180° domain walls forms several domain pairs, while the 90° domain walls enable transitions between different domain pairs (yellow/green pairs and blue/red pairs), ultimately resulting in an intertwined domain pairs structure at region 1.

The region 2 features four equivalent domain variants arranged in an anticlockwise vortex-like structure. As illustrated in Figure 3d, this region 2 contains two 90n domain walls and two 90c domain walls, and these four domain walls are symmetrically arranged around a central vortex domain structure. There are no charged 180° domain walls with out-of-plane shuffle in this region, and only four 90° domain walls contribute to the formation of the vortex topological structure. Figure 3e illustrates the domain walls structure of 90n and 90c at region 3, highlighting the distinctly different local bonding configurations. In the 90n domain walls, the Bi-Bi bonds are perpendicular to the domain wall, whereas in the 90c domain walls, the bonds are parallel to the domain wall. The top view illustrates the polarization switch around 90° domain walls, transitioning from $(0, -P_s)$ to $(-P_s, 0)$, and then to $(0, +P_s)$, achieved through a bond-breaking mechanism. DFT calculations show moderate energy barriers of these transitions (Figure S13, Supporting Information).

Here, we repeated the simulations of cooling several times and obtained various multi-domain configurations. Most of the time, the multi-domain structures have four equivalent domain variants which contain both 180° and 90° domain walls, though occasionally pure 180c domains (Figure

S14, Supporting Information) and pure 90° domains (Figure 2a) were obtained. Some examples of multi-domain are illustrated in the Figure S15a of the Supporting Information, all of which contain both 180° and 90° domain walls. Taking domain 1 as an example, when a compressive strain ($\varepsilon_x=$ -3.08%) is applied along the *x*-direction, it transforms into the standard checkerboard domain structure, indicating that it is essentially a distorted checkerboard domain (Figure S15b, Supporting Information). The checkerboard domain is a commonly existing configuration.

**The evolution of domain-pair intertwined checkerboard domains upon strain**

We discovered that this topological structure shows excellent stability against the applied deformation. We performed a quasi-static tensile test along the *x* direction on the checkerboard domain at 10 K. Figure 4a and 4b illustrate the evolution of stress ($\sigma_x$) and potential energy as the functions of strain. Each point on the curves corresponds to a thermodynamic equilibrium state in the MD simulations at a given strain level. We selected a site of typical configurations (letters a-l) during the strain process. The stress and energy curves exhibit three distinctive stages: domain wall motion ('a-d'), 180° domain wall reaction ('e-h'), and 90° domain wall reaction ('i-l'). These three stages are respectively highlighted in green, red, and yellow boxes, illustrated in Figure 4b.

Figure 4c illustrates the motion of 180° and 90° domain walls and the stable topological structure in stage I ($\varepsilon_x$ = 0 to 4.70%). To guide the visualization, the 180° domain walls are highlighted as solid black lines. The black arrows indicate the movement of the 90° domains. Due to the coupling of strain and polarization, the nano-domain pairs with the ($\pm P_s$, 0) polarizations gradually expand as the strain increases up to $\varepsilon_x$ = 4.70%. As shown in insets 'a-d', the domain pair (bule and red colors) expand through the motion of 90° domain walls, also driving the 180° domain walls bending along the *y* direction. The dominate domain pairs switch from (0, $\pm P_s$) to ($\pm P_s$, 0) solely through domain wall motion, without nucleation, while maintain the same topological structure. Moreover, the entire stage I of domain wall motion and topological preservation requires a very small stress ($\approx$ 5 MPa) to drive, with only a slight increase in potential energy, and can be fully recovered by applying an opposite strain. We also performed a similar strain-induced 90° domain switching on a single domain. The results show that the 90° domain switching in the single domain involves a distinct nucleation process and requires significantly higher stress ($\approx$ 251 MPa) to initiate. There is a noticeable yielding behavior (corresponds to the nucleation process) observed in both the energy and stress curves (Figure S16, Supporting Information).

Figure 4d illustrates the typical configurations of the 180° domain wall reaction, occurring during the stage II ($\varepsilon_x$= 4.70% to 8.32%). The black solid boxes (1, 2, and 3) highlight the evolution of a specific region. As shown in inset 'e' ($\varepsilon_x$= 5.12%), the 180° domain walls bend and converge in the *y* direction at that specific region. Figure 4e provides a zoom-in view of that region, which contains multiple domain walls (180c, 90n, and 90c). This structure prevents the further motion of domain walls, resulting a notable increase in both stress and energy. As the strain increases to $\varepsilon_x$= 6.18%, two 180° domain walls cross in the previous region, forming an interconnected domain wall network. With further strain increase ($\varepsilon_x$= 7.89%), the domain wall network splits along the *x* direction, ultimately resulting in two new 180° domain walls (inset 'h') that are perpendicular to the initial 180° domain walls. Meanwhile, during the 180° domain wall reaction, the potential energy and stress gradually increase smoothly. After fully optimizing the structure of inset 'h' ($\varepsilon_x$= 8.32%), the multi-domain configuration reveals another standard checkerboard domain (Figure 4f), which is equivalent to the initial checkerboard domain but rotated by 90 degrees.

Figure 4g illustrates the typical configurations of 90° domain wall reaction, occurring during the stage III ($\varepsilon_x$ = 8.32% to 9.41%). As shown in inset 'i' ($\varepsilon_x$ = 8.54%), the 90° domain walls split at the intersection with the 180° domain walls. With further strain increase, the two oppositely configured 90° domain walls move closer together (inset 'j' and 'k'), gradually annihilating and ultimately disappearing completely. Eventually, the checkerboard domain transforms into a pure 180° domain structure (inset 'l'). Additionally, a noticeable yielding behavior is observed in both the energy and stress curves, which can be attributed to the reduction in domain wall area during the 90° domain wall reaction. For comparison, we performed a similar tensile test along the *y* direction on the checkerboard domain to observe the evolution of the domain structure (Figure S17, Supporting Information). Since the initial dominant domain pairs align with the tensile direction, the domain wall motion or subsequent 180° domain wall reactions are not observed. Instead, the system directly underwent a 90° domain wall reaction, resulting in a pure 180° domain structure.

**Discussion and Conclusion**

Using machine-learning potential and molecular dynamics (MD) simulations[34–37], we investigated the finite-temperature dynamics of domain walls and discovered a domain-pair intertwined topological domain structure in Bi monolayer. First, Bi monolayer has a second-order ferroelectric phase transition, which is quite different from other 2D ferroelectric materials like GeSe[38,39], despite their similar structure. The GeSe has a first-order ferroelectric transition with discontinuities in lattice constant.

Finite-temperature dynamics reveal that 180° domain walls exhibit significantly higher mobility compared to the 90° domain walls at 300 K, which may related to the unique shuffling polarization switching mechanism in 180° domain walls. This mechanism is characterized by the out-of-plane shuffle of Bi atoms without bond breaking. As the shuffle occurs, the intrinsic buckling ($d_z$) reverses, leading to charge redistribution between Bi atoms, where Bi anions become cations and vice versa, ultimately reversing the polarization. In contrast, the domain switching in 90° domain walls is driven by traditional in-plane ionic movements, similar to GeSe monolayer[38], rather than charge property reversal. The single-element nature allows for easier reversal of charge properties, while achieving such charge reversal in compounds (such as GeSe) with elements of differing electronegativities is more challenging.

After cooling the sample from 500 K to 300 K, we obtained a domain-pair intertwined checkerboard domain structure, coexisting of 180° and 90° domain walls. The head-to-head configurations around charged 180° domain walls (with the lowest energy) give rise to two groups of domain pairs, with the 90° domain walls appearing between different domain pairs. These two groups of domain pairs form a vortex-like topological structure in region 2. This checkerboard domain maintains a stable topological structure within the strain range ($\varepsilon_x$ = 0 to 4.70%), and the domain motion without nucleation allows the minimal stress (≈ 5 MPa) and a slight energy increase. Additionally, with further strain increase, there are rich domain wall reactions and the resulted 180c domain structure.

In conclusion, we investigated the finite-temperature dynamics of domain walls and discovered a domain-pair intertwined topological domain in Bi monolayer. Our findings reveal two distinct formation mechanisms for 180° and 90° domain walls. Furthermore, we discovered a checkerboard domain characterized by intertwined domain pairs and coexistence of 180° and 90° domain walls, demonstrating a stable topological structure within a small strain range. Meantime, it exhibits rich domain wall reactions and domain structure evolution, which may hold significant promise for applications[10,40] in advanced data storage, neuromorphic computing, quantum computing, energy-efficient devices, flexible electronics, and next-generation sensors.

## Methods

**DFT.** Our density functional theory (DFT) calculations for the training database and domain wall energies were performed using the VASP software[41,42]. We employed the Perdew-Burke-Ernzerhof (PBE) exchange-correlation functional[43] and the projected augmented plane-wave (PAW) method[44]. The Bi monolayer was modeled in the *x-y* plane with a vacuum spacing of approximately 20 Å along the *z*-axis to prevent artificial interactions. The valence configuration was treated as $6s^2 6p^3 5d^{10}$. An energy cutoff of 500 eV was applied, and the Brillouin zone was sampled in k-space with a minimum grid spacing of 0.2 Å$^{-1}$. The crystal structure optimization was considered converged when the energy difference between consecutive steps was less than $10^{-6}$ eV, and the maximum force on any atom was below 0.001 eV/Å. Some representative datasets capturing physical phenomena were generated using *ab initio* molecular dynamics (AIMD) simulation[45]. These AIMD simulations were performed in the NVT ensemble with a Nosé–Hoover thermostat, using a timestep of 3 fs and running up to at least 3 ps.

**Machine learning potential.** We used the Deep Potential Generator[46] (DPGEN) scheme to develop a deep neural network-based (DNN) potential of Bi monolayer. This machine learning potential was directly learned from an accurate reference database of DFT calculations. The DPGEN framework features an activate learning[47] procedure organized in a closed loop consisting of three steps: training, exploration, and labeling. In the training step, four models are trained using the same database of DFT energies and forces, but with different initial values of hyperparameters of the deep neural networks (DNNs). These four models were employed to perform MD simulations in the next exploration step, with the model deviation serving as an indicator of sampling importance. If the configuration is well represented by the current training database, the four DP models are expected to demonstrate nearly the same predictive accuracy. Some poorly represented configurations will be labeled for additional DFT calculations and added to the database for training in the next loop. The DPGEN scheme allows for automatic, iterative, and efficient updates to the training database, significantly reducing the impact of the initial database construction on the force field development and enhancing the reliability of the models.

The training database consists of structures generated by randomly perturbing ground-state *Pmn2₁* structures. For DFT calculations, we used a 3×3×1 supercell with 36 atoms, with the details described in the DFT section above. During the exploration step, NPT simulations were run at temperatures ranging from 50 to 800 K and pressures from -100 to 1000 bar to sample the configuration space. Additionally, some representative datasets capturing physical phenomena such as domain switching, 90° domain walls, uniaxial strain, and the antiferroelectric phase (as previously reported) were generated using ab initio molecular dynamics (AIMD) simulation. These datasets can be used to both test the model's accuracy in describing these physical phenomena and further enhance the model's predictive accuracy for these processes.

**MD:** Classic molecular dynamics (MD) simulations were performed using a timestep of 1 fs, with periodic boundary conditions applied in all three dimensions. All simulations were conducted at various temperatures and a constant pressure of 1 atmosphere (1 bar). The Nosé-Hoover thermostat and barostat were employed to control temperature and pressure under the NPT ensemble. Atomic configurations of Bi monolayer, containing 40000 atoms, were visualized using the VESTA software. Typical polarization configurations were analyzed and visualized with Python code based on the atomistic order parameter ($\vec{R}$) and the color map in figure 1b.

**Randomly polarized structure:** The randomly polarized structure was generated using Python code, with polarization vectors randomly oriented in all directions.

**Acknowledgment**

The authors gratefully acknowledge the support of NSFC (Grant No. 11974269), the support of the Key Research and Development Program of Shaanxi (Grant No. 2023-YBGY-480), and the 111 Projects 2.0 (Grant No. BP0618008). J. Z. L. acknowledges the support from ARC discovery projects (Grant No. DP210103888) and HPC from National Computational Infrastructure from Australia. This work is supported by the Bohrium Cloud Platform of DP Technology. The authors also thank F. Yang and X. D. Zhang at the Network Information Center of Xi'an Jiaotong University for supporting the HPC platform.

Figure 1

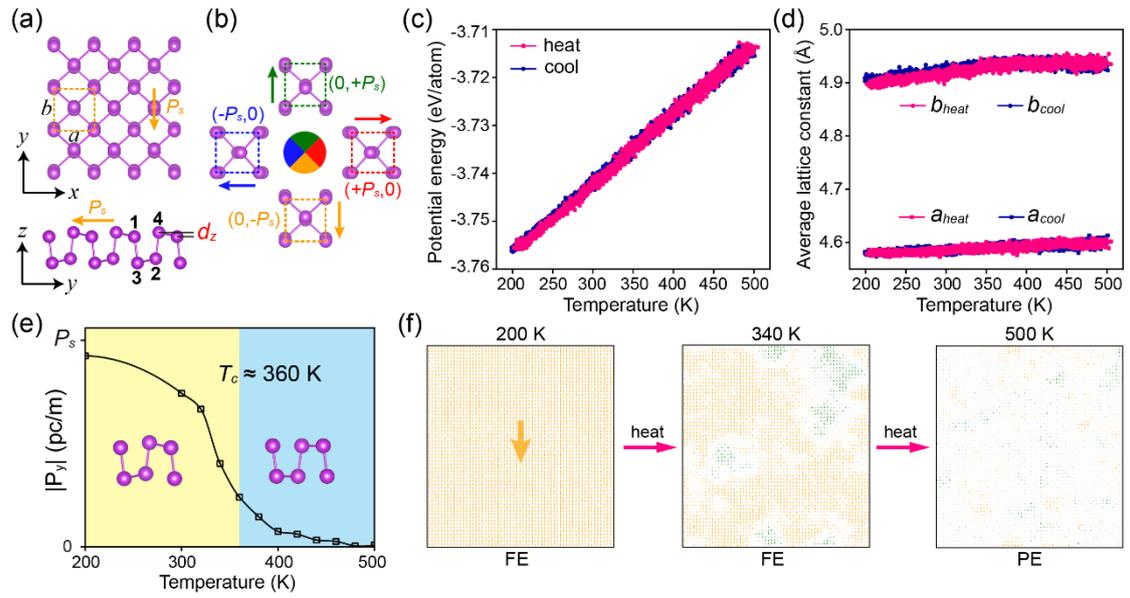

**Figure 1: Temperature induced second-order ferroelectric phase transition in Bi monolayer.**
(a) Crystal structure of Bi monolayer. The vacuum layer was set in the *z* direction. The $d_z$ represent the *z*-direction displacement of neighboring Bi atoms. (b) Four equivalent domain variants with different intrinsic polarization ($P_s$), represented by $(0, +P_s)$, $(0, -P_s)$, $(+P_s, 0)$, and $(-P_s, 0)$ vectors. The arrows with different colors also indicate the directions of polarization. (c) The evolution of potential energy as the function of temperature during heating and cooling. (d) The evolution of average lattice constant as the function of temperature during heating and cooling. (e) The evolution of polarization as the function of temperature during isothermal simulation. Yellow and blue regions indicate the temperature ranges for the ferroelectric and paraelectric phase, along with the corresponding average atomic structures. (f) Typical atomic configurations at specific temperatures (200 K, 340 K, and 500 K). Each arrow represents one unit cell with four Bi atoms. The colors indicate the direction of polarization, with the color map shown in figure 1b.

**Figure 2**

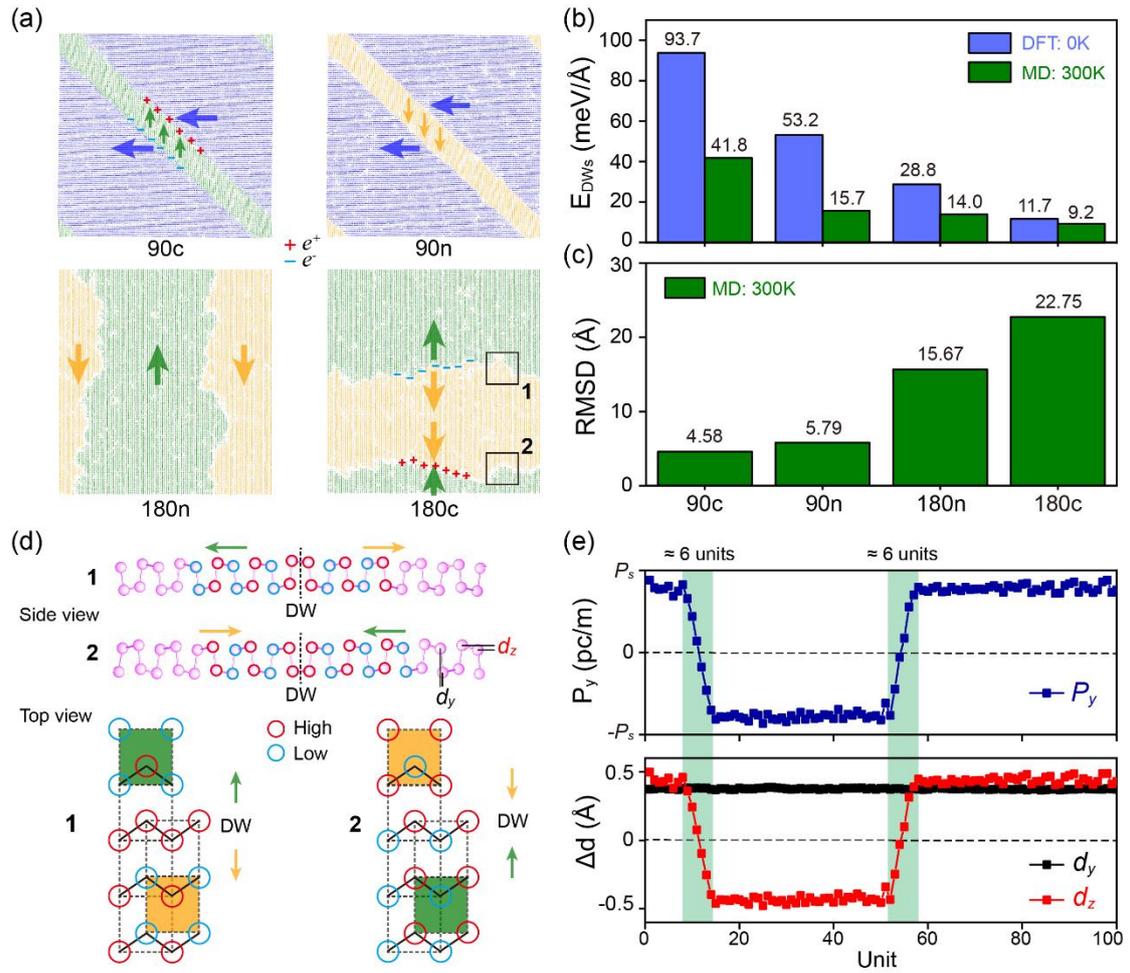

**Figure 2: Finite-temperature dynamics of four typical domain walls.** (a) Atomic structures of four typical domain walls (180n, 180c, 90n, and 90c) at 300 K. The arrows represent the direction of polarization but don't indicate its magnitude. Different from the neutral domain walls (180n, 90n), bound charges ($e^+$ and $e^-$) at the charged domain walls (180c, 90c) are indicated by positive (red) and negative signs (blue), respectively. (b) Comparison of domain wall energies for four typical domain walls at 300 K. The blue and green bars represent the results from DFT and MD simulations, respectively. (c) Comparison of domain wall mobilities for four typical domain walls at 300 K. The root mean square displacement (RMSD) was used to quantify the domain walls mobility. (d) Atomic structure of charged 180° domain walls. Red and blue circles indicate the Bi atoms with different distortions (high and low) in the $z$ direction. The arrows and dashed lines indicate the polarization and the domain walls. The top view only shows the higher sub-layer Bi atoms and the chemical bonds (Solid black line). (e) Quantitative analysis of atomic structure and polarization switching. The definitions of $d_z$ and $d_y$ are shown in the side view of figure 2d. The domain walls regions are highlighted with green boxes.

**Figure 3**

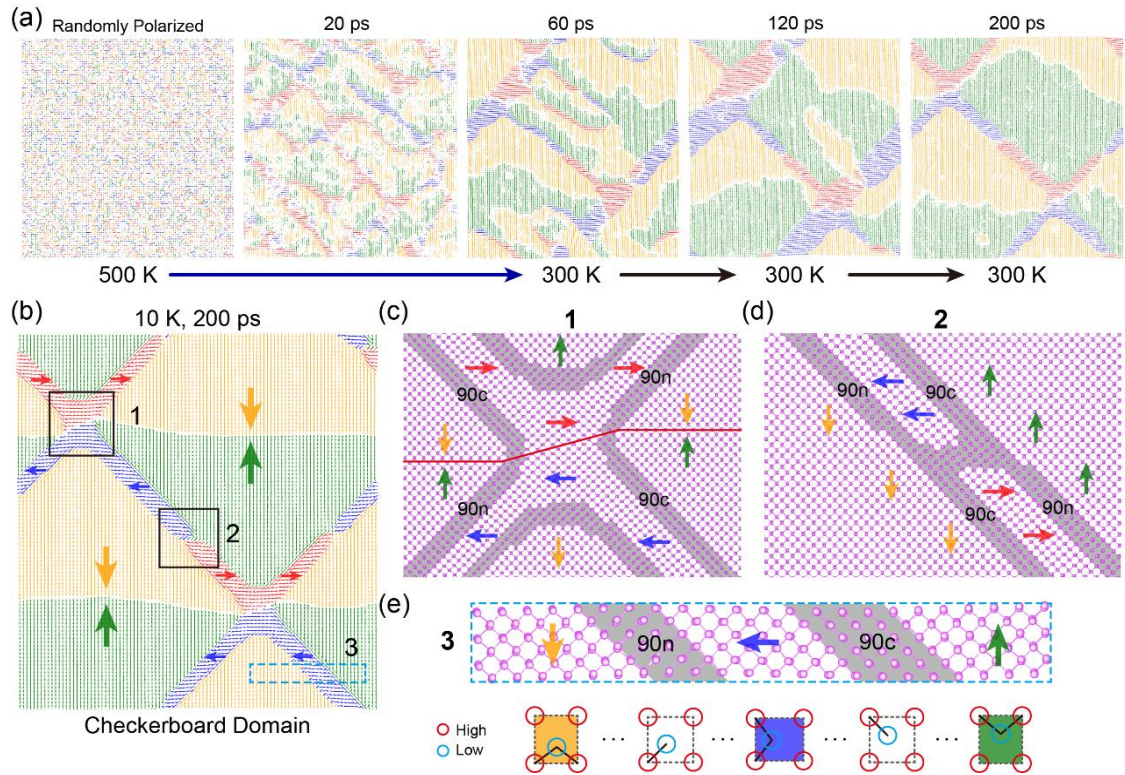

**Figure 3: Checkerboard domain with coexisting 180° and 90° domain walls.** (a) Formation and evolution of the checkerboard domain during cooling and isothermal process. The polarization distribution of initial artificial structure is completely random. The colors indicate the direction of polarization, as depicted in the color map in figure 1b. (b) The atomic structure and polarization distribution of checkerboard domain at 10 K. The arrows and colors indicate the direction of polarization, with the color map shown in figure 1b. The black solid boxes (1, 2) highlight two specific regions where multiple domains intersect, while the blue dashed box marks the third region featuring two distinct 90° domain walls (90c and 90n). (c) The atomic structure of the region 1, where 180° and 90° domain walls intersect. The gray areas highlight the 90° domain walls (with broken bonds), while the red solid lines indicate charged 180° domain walls (without broken bonds). (d) The atomic structure of the region 2, featuring a vortex-like domain structure. (e) The atomic structure of the region 3. The top view only shows the higher sub-layer Bi atoms and the chemical bonds (Solid black line). The gray areas highlight the broken bonds in 90° domain walls.

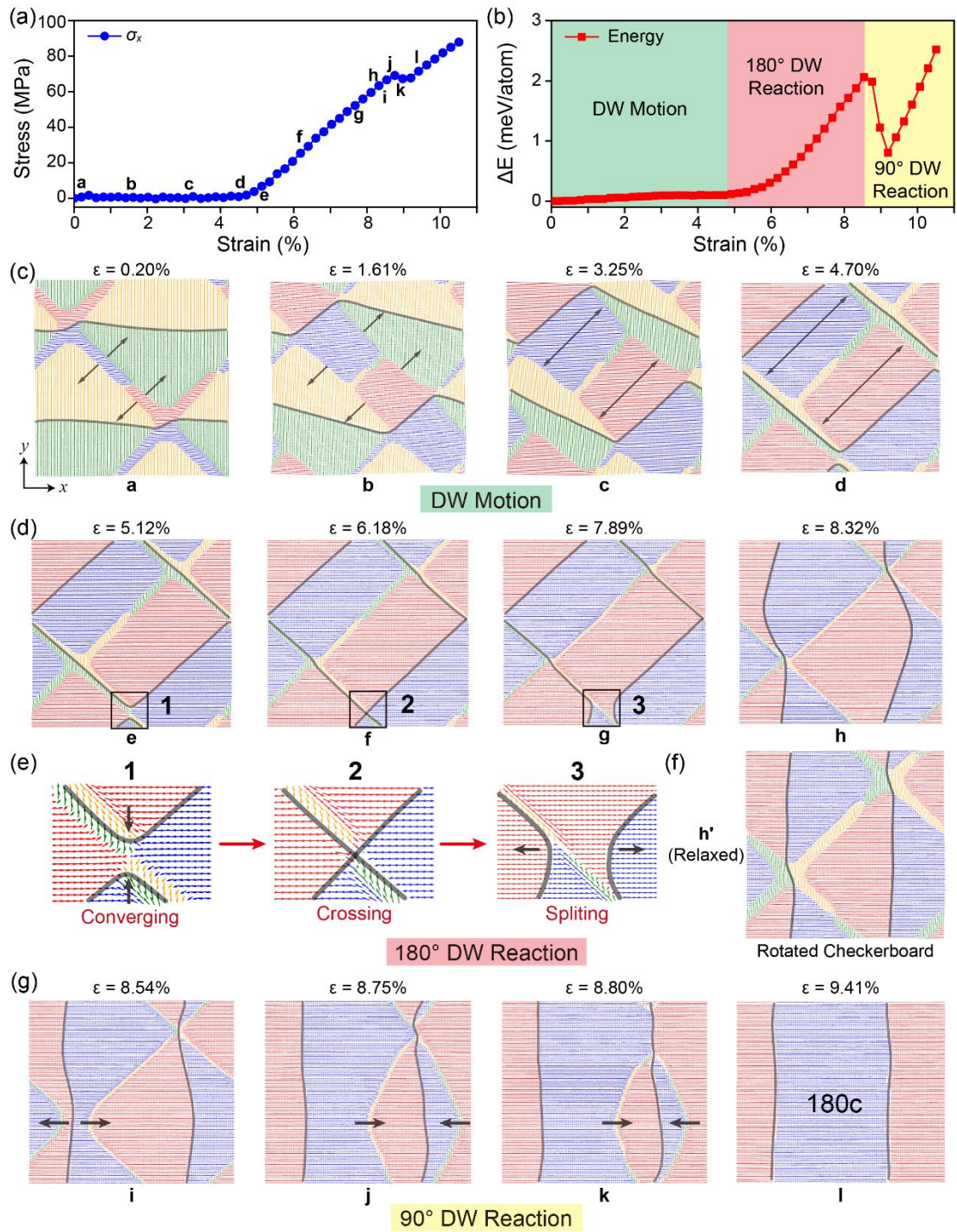

**Figure 4: The evolution of checkerboard domain as the function of strain.** (a) The evolution of stress ($\sigma_x$) as the function of strain. Letters (a-l) represent typical configurations during strain induced domain switching process. (b) The evolution of energy as the function of strain. The green, red, and yellow boxes represent the stages of domain wall (DW) motion, 180° DW reaction, and 90° DW reaction, respectively. (c) The motion of domain wall and stable topological structure occur in the first stage ($\varepsilon_x$ = 0 to 4.70%). The black arrows indicate the direction of domain wall movement, and the 180° DWs are highlighted with solid black lines. (d) The reaction of 180° DW and the resulting rotated checkerboard domain occur during the second stage ($\varepsilon_x$ = 4.70% to 8.32%). The black solid boxes (1, 2, and 3) highlighted the specific region of the reaction. (e) The evolution of the typical region, showing the details of 180° DW reaction. Numbers (1, 2, and 3) represent three typical atomic configurations during the reaction. (f) The relaxed structure of configuration 'h', which illustrates the configuration after the 180° DW reaction. (g) The reaction of the 90° DW and the resulting charged 180° domain occur during the third stage ($\varepsilon_x$= 8.32% to 9.41%).

# Supplemental Information

**TABLE S1**: Comparison of DFT and DP results for the crystal structure ($a$, $b$, and $d_z$) and energy of the Bi monolayer. The parameter $d_z$ represents the intrinsic buckling, as shown in Figure 1a.

|     | $a$ (Å) | $b$ (Å) | $d_z$ (Å) | Energy (eV/atom) |
| --- | --- | --- | --- | --- |
| DFT | 4.57 | 4.85 | 0.49 | -3.78294 |
| DP  | 4.58 | 4.88 | 0.47 | -3.78289 |

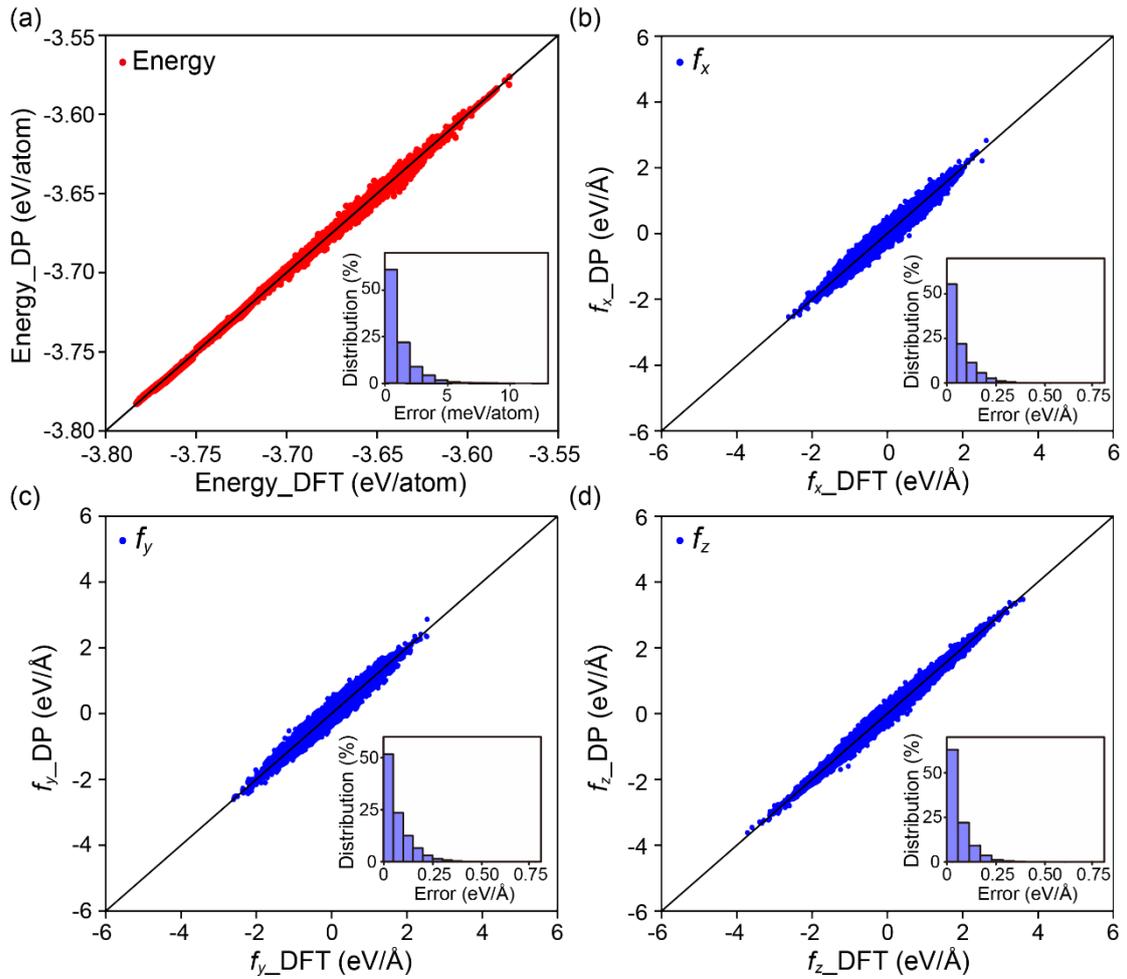

**Figure S1:** Comparison of DFT and DP results for (a) energy per atom and [(b)-(d)] atomic forces for configurations in the final training database. The insets display the distributions of the absolute errors, while the diagonal lines indicate the perfect correlations between the corresponding variables.

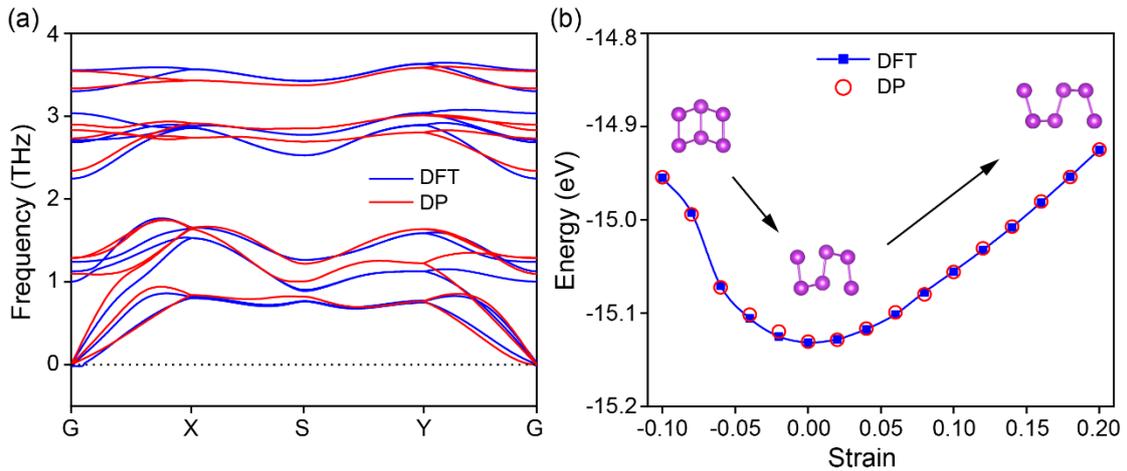

**Figure S2:** Comparison of DFT and DP results for (a) phonon spectrum and (b) energy-strain curve. The insets of (b) display three distinct atomic configurations under different strains, demonstrating the significant effect of strain on the Bi monolayer.

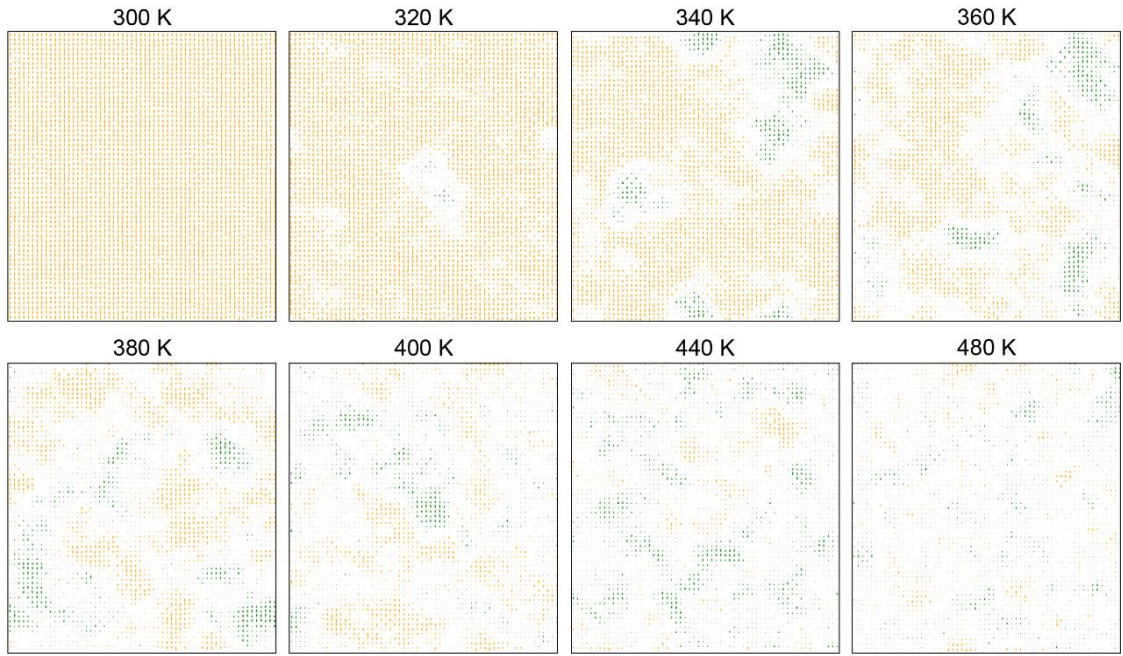

**Figure S3:** The evolution of atomic configurations as the function of temperature, during 200 ps isothermal simulations of a single domain at various temperatures. Each inset illustrates the average position over the last 10 ps, showing the gradual disappearance of long-range polarization.

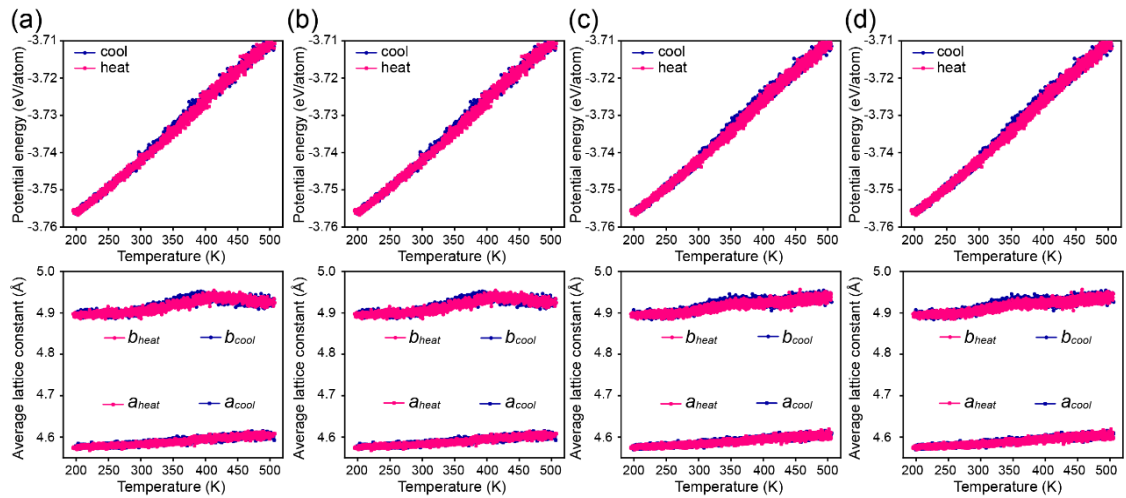

**Figure S4:** The evolution of potential energy and average lattice constant as the function of temperature during heating and cooling cycles. Insets (a)-(d) represent four heating and cooling cycles, showing the similar characteristics.

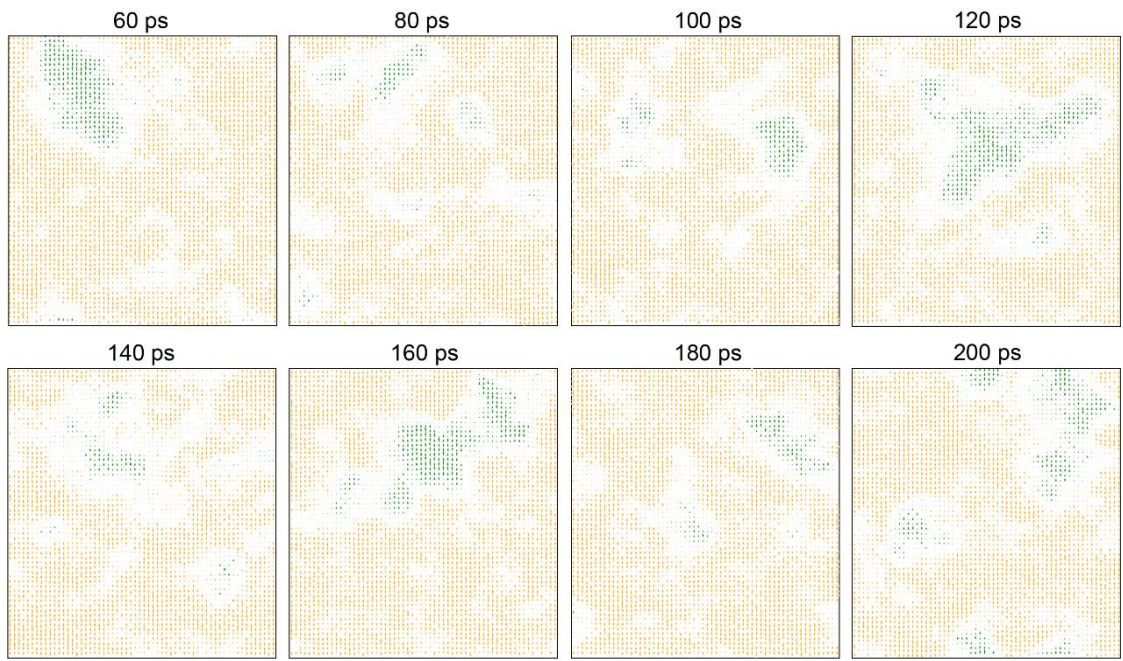

**Figure S5:** The evolution of atomic configurations as the function of time at 340 K. The nano-regions with different polarization orientations are not fixed while randomly triggered over time, leading to the reduction in total polarization.

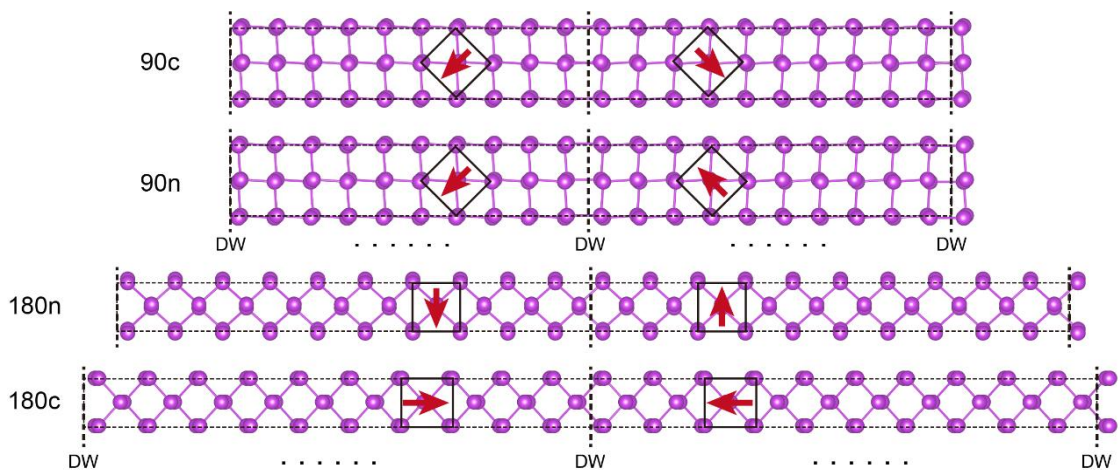

**Figure S6:** DFT models of four typical domain walls: charged 90° (90c), neutral 90° (90n), neutral 180° (180n), and charged 180° (180c) domain walls. The dashed black boxes represent the supercell of the domain wall models, while the solid black boxes indicate the unit cell. The red arrows show the polarization direction of each domain, and the positions of domain walls are highlighted by dashed black lines.

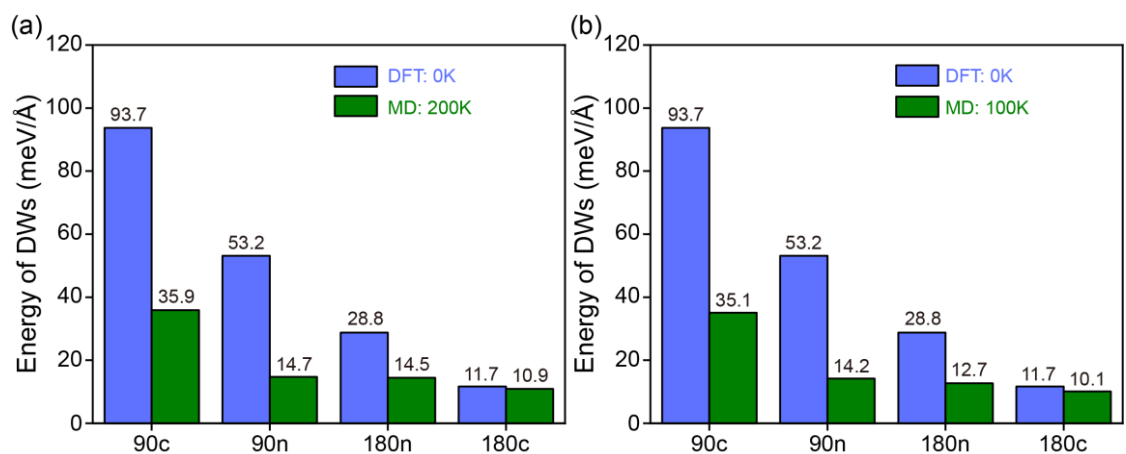

**Figure S7:** Comparison of domain wall energies for four typical domain walls at (a) 200 K and (b) 100 K. The blue bars represent the results from DFT simulations, while the green bars indicate the results from MD simulations. At both 200 K and 100 K, the energy relationship is observed as: $E_{DWs}(90c) > E_{DWs}(90n) > E_{DWs}(180n) > E_{DWs}(180c)$.

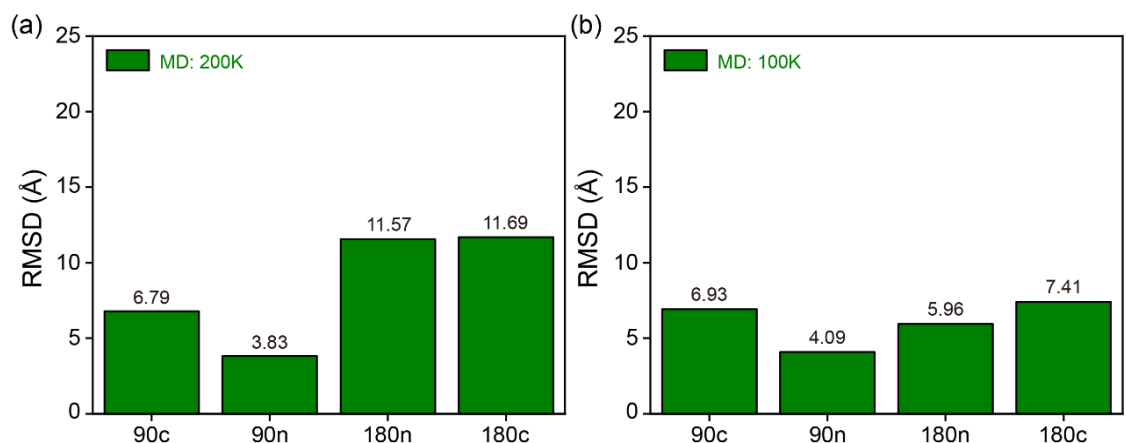

**Figure S8:** Comparison of domain wall mobilities for four typical domain walls at (a) 200 K and (b) 100 K. At 200 K, the 180° domain walls exhibit better mobility, while this difference disappears as the temperature decreases to 100 K.

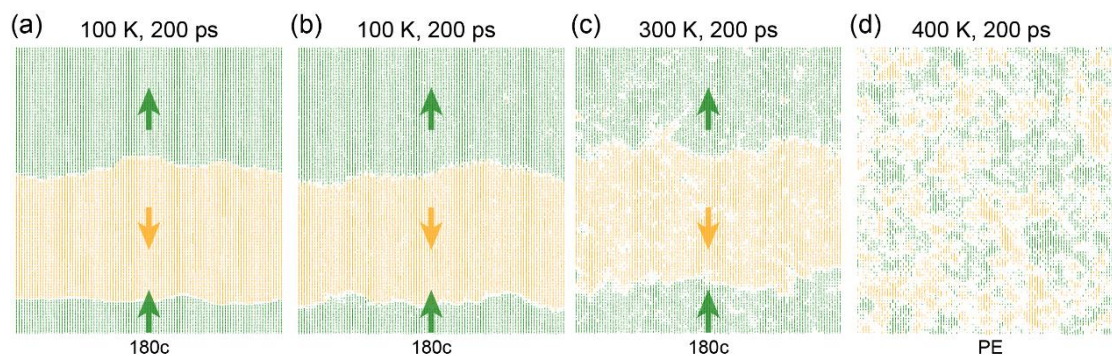

**Figure S9:** Atomic configurations of charged 180° domain walls under various temperatures (100 K, 200 K, 300 K, and 400 K). The results illustrate the domain wall mobility significantly increases in high temperature, resulting in structural irregularities in the domain walls.

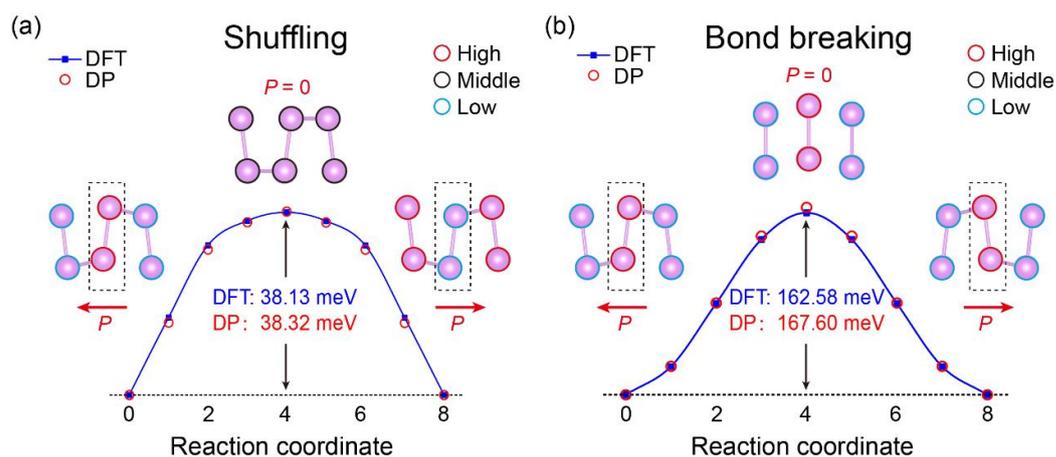

**Figure S10:** Comparison of the DFT and DP results for energy profiles of polarization reversal pathways in Bi monolayer. (a) Switching polarization through *z*-direction shuffling without bond breaking. Different vertical Bi pairs are highlighted with red, black, and blue circles. The black dashed boxes illustrate the vertical Bi pairs shuffle from 'High' to 'Low'. (b) Switching polarization through bond breaking mechanism. The black dashed boxes highlight the change of bonding, but the vertical Bi pairs are always 'High'.

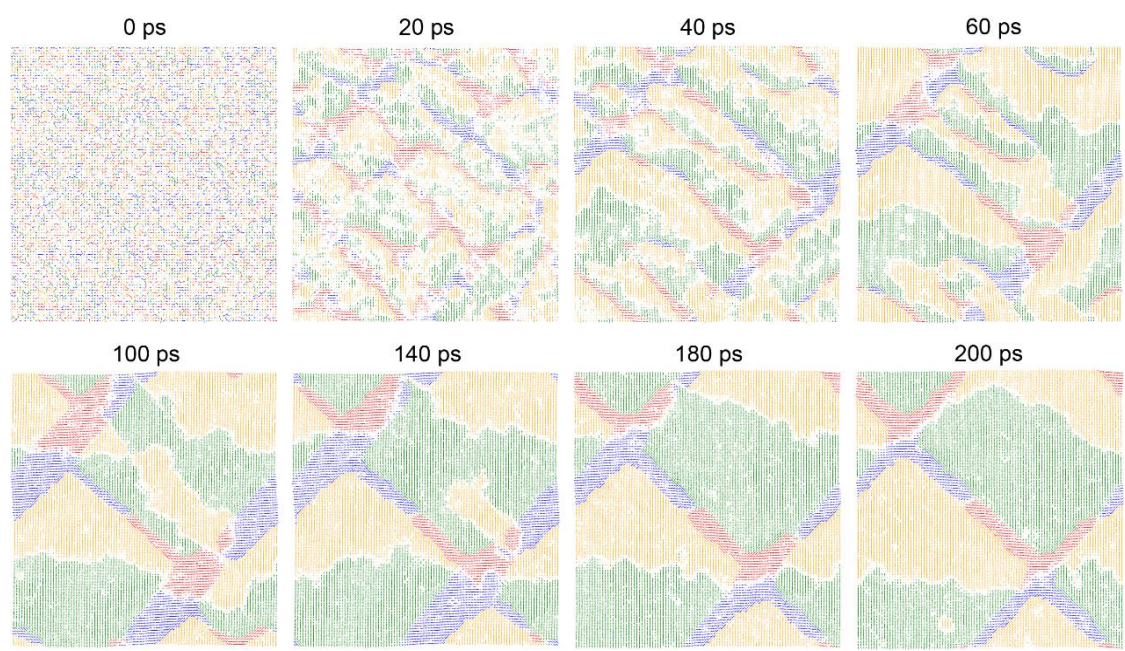

**Figure S11:** Formation and evolution of the checkerboard domain during cooling from 500 K (0 ps) to 300 K (60 ps) and the additional isothermal process. The initial artificial structure at 0 ps exhibits completely random polarization, with colors representing the polarization directions, as shown in the color map in figure 1b.

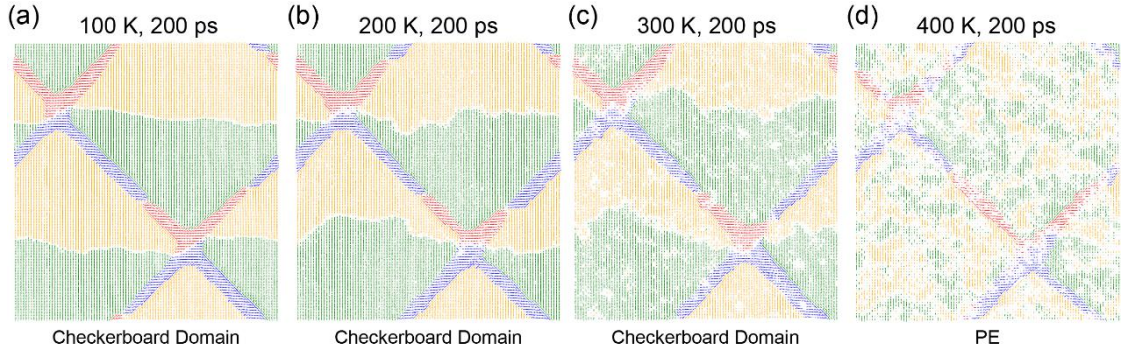

**Figure S12:** Atomic configurations of checkerboard domain under various temperatures (100 K, 200 K, 300 K, and 400 K). The results illustrate the stability of checkerboard domain at (a) 100 K, (b) 200 K and (c) 300 K, while it transformed into the paraelectric phase at (d) 400 K.

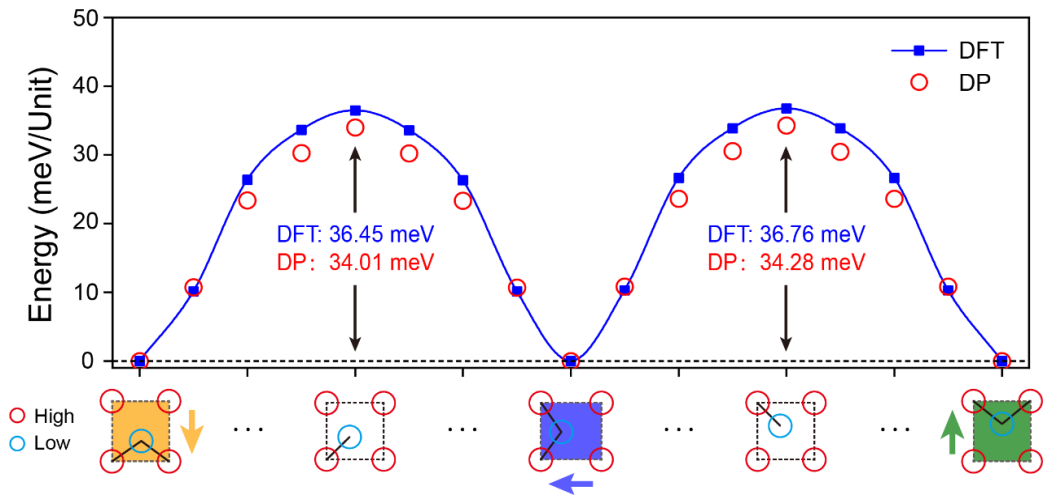

**Figure S13:** Comparison of the DFT and DP results for energy profiles of polarization switching pathways in Bi monolayer, transitioning from $(0, -P_s)$ to $(-P_s, 0)$, and then to $(0, +P_s)$. The typical atomic structures during the transition are illustrated at the bottom using different colors, as shown in the color map in figure 1b. The top view only highlights the upper sub-layer Bi atoms and the chemical bonds, indicated by solid black lines.

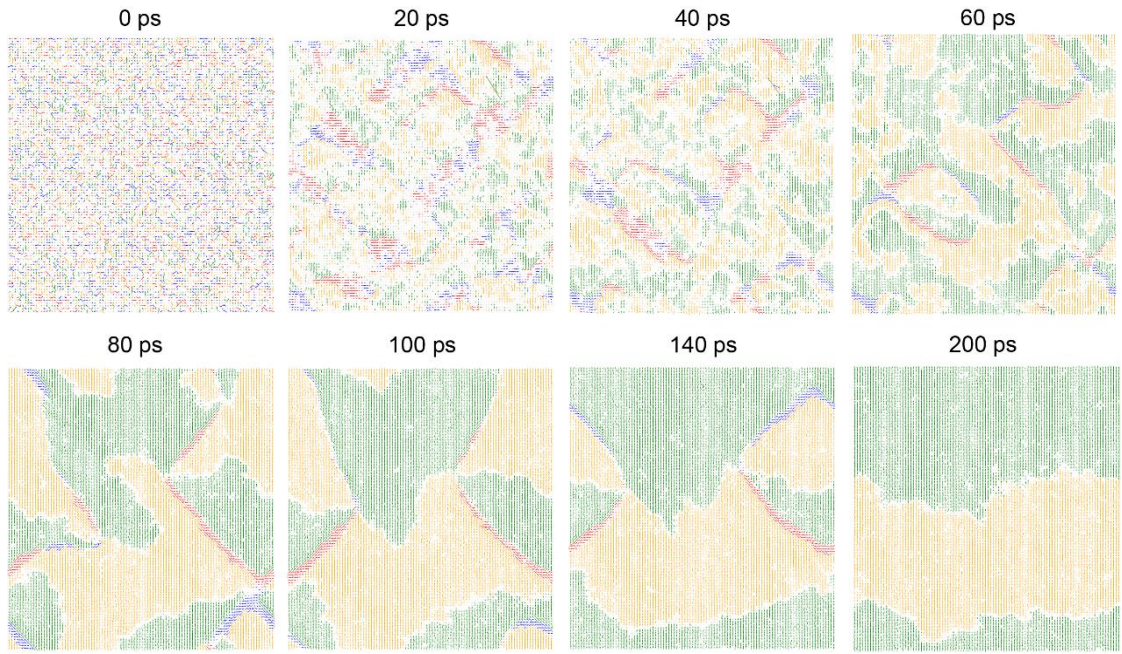

**Figure S14:** Formation and evolution of the charged 180° domain during cooling from 500 K (0 ps) to 300 K (60 ps) and isothermal (300 K) process. The initial artificial structure (0 ps) is completely randomly polarized. The colors indicate the direction of polarization, as depicted in the color map in figure 1b.

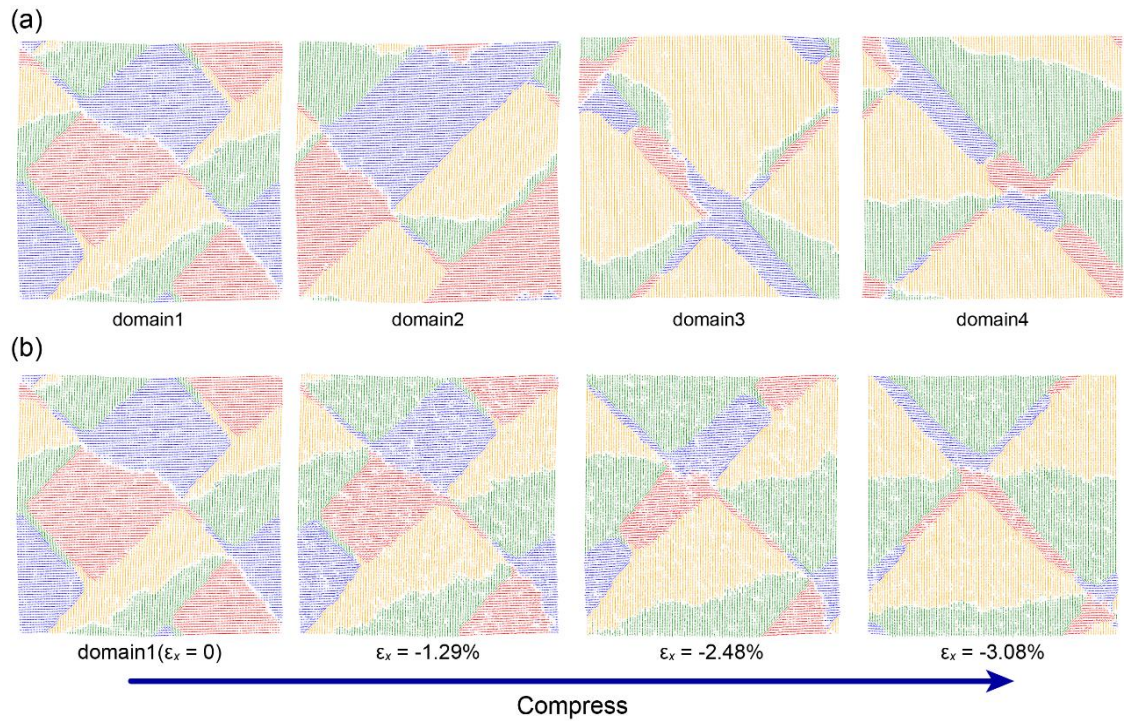

**Figure S15:** Various multi-domain configurations with four equivalent domain variants are obtained from the cooling simulations. (a) Examples of multi-domain configurations observed in the MD simulations, each containing both 180° and 90° domain walls. (b) The evolution of Domain 1 configurations as a function of compressive strain, illustrating typical configurations during the strain-induced domain switching process.

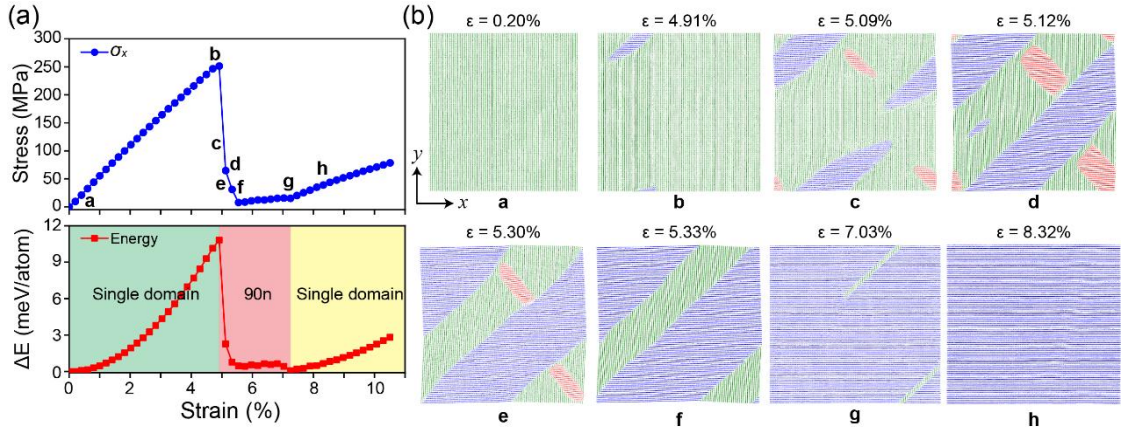

**Figure S16:** The evolution of a single-domain as a function of $x$-direction strain. (a) The evolution of stress ($\sigma_x$) and energy as the function of strain, revealing three distinct stages: single domain, neutral 90° domain (90n), and single domain. Each point on the curves corresponds to a stable state at a given strain level. Letters (a-h) correspond to typical configurations during the strain-induced domain switching process. (b) Typical configurations during the evolution of single domain, illustrating the transition from a $(0, +P_s)$ single domain to a 90° domain, and finally to a $(-P_s, 0)$ single domain.

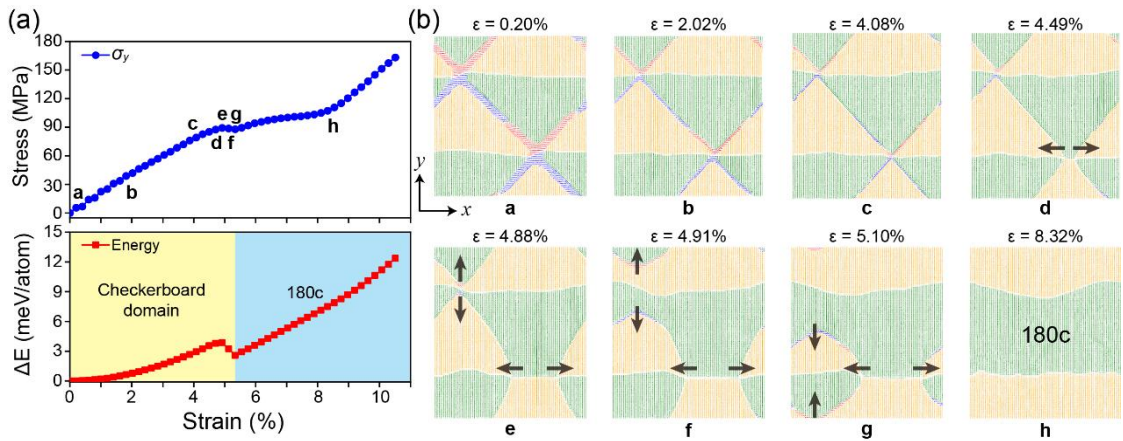

**Figure S17:** The evolution of checkerboard domain configurations as a function of $y$-direction strain. (a) The evolution of stress ($\sigma_y$) and energy as the function of strain, revealing two distinct stages: checkerboard domain and charged 180° domain (180c). Each point on the curves corresponds to a stable state at a given strain level. Letters (a-h) represent typical configurations during strain induced domain switching process. (b) Typical configurations during evolution of checkerboard domain. The black arrows indicate the direction of domain wall movement.